\documentclass[preprint,aps,aps,showpacs,onecolumn,superscriptaddress]{revtex4}
\usepackage{amsfonts}
\usepackage{amsmath}
\usepackage{amssymb}
\usepackage{graphicx}
\usepackage{txfonts}

\begin{document}

\title{Ultimate phase estimation in a squeezed-state interferometer using
photon counters with a finite number resolution}
\author{P. Liu}
\affiliation{Department of Physics, Beijing Jiaotong University, Beijing 100044, China}
\author{G. R. Jin}
\email{grjin@bjtu.edu.cn}
\affiliation{Department of Physics, Beijing Jiaotong University, Beijing 100044, China}
\date{\today }

\begin{abstract}
Photon counting measurement has been regarded as the optimal measurement
scheme for phase estimation in the squeezed-state interferometry, since the
classical Fisher information equals to the quantum Fisher information and
scales as $\bar{n}^2$ for given input number of photons $\bar{n}$. However,
it requires photon-number-resolving detectors with a large enough resolution
threshold. Here we show that a collection of $N$-photon detection events for
$N$ up to the resolution threshold $\sim \bar{n}$ can result in the ultimate
estimation precision beyond the shot-noise limit. An analytical formula has
been derived to obtain the best scaling of the Fisher information.
\end{abstract}

\maketitle

\section{Introduction}

Quantum phase estimation through a two-path interferometer (e.g., the widely
adopted Mach-Zehnder interferometer) is well-known inferred from the
intensity difference between the two output ports. With a coherent-state
light input, the Cram\'{e}r-Rao lower bound of phase sensitivity can only
reach the shot-noise (or classical) limit~\cite%
{Helstrom,Kay,Braunstein,Luo,Smerzi09,Giovannetti}, $\delta \varphi _{%
\mathrm{CRB}}=1/\sqrt{F(\varphi )}\sim 1/\sqrt{\bar{n}}$, where $F(\varphi
)\sim O(\bar{n})$ denotes the classical Fisher information and $\bar{n}$ is
the mean photon number. To beat the classical limit, Caves~\cite{Caves}
proposed a squeezed-state interferometer by feeding a coherent state $%
|\alpha \rangle $ into one port and a squeezed vacuum $|\xi \rangle $ into
the other port, as illustrated by the inset of Fig.~\ref{fig1}, which is of
particular interest for high-precision gravitational waves detection~\cite{Caves,Aasi} and new generation of fountain clocks based on atomic squeezed vacuum~\cite{Kruse,Peise}.

Theoretically, Pezz\'{e} and Smerzi~\cite{Smerzi08} have shown that photon-counting measurement is optimal in the squeezed-state interferometer, since the classical Fisher
information (CFI) equals to the quantum Fisher information (QFI) and scales
as $\bar{n}^{2}$, leading to the ultimate precision in the Heisenberg limit $%
\delta \varphi _{\mathrm{CRB}}\sim 1/\bar{n}$. Recently, the phase-matching
condition that maximizes the QFI has been investigated~\cite{Liu}. Lang and
Caves~\cite{Lang} proved that under a constraint on $\bar{n}$, if a
coherent-state light is fed from one input port, then the squeezed vacuum is
the optimal state from the second port.

The theoretical bound in the phase estimation~\cite{Smerzi08,Liu,Lang} has
been derived by assuming photon-number-resolving detectors (PNRDs) with a
exactly perfect number resolution~\cite{Seshadreesan}. However, the best
detector up to date can only resolve the number of photons up to $4$~\cite{Smerzi07,Kardynal}. Such a resolution threshold is large enough to realize
coherent-state light interferometry with a low brightness input $\bar{n}%
\simeq 1$~\cite{Smerzi07}. To achieve a high-precision quantum metrology,
nonclassical resource with large number of particles is one of the most
needed~\cite{Slusher1985,LAWu1986,Slusher1987,Breitenbach97,Vahlbruch08,Dowling15,Ma,YRZhang,Toth,Tan,Matthews,Luca}. For an optical phase estimation, it also requires the interferometer with
a low photon loss~\cite{Donner,Joo,Zhang,Knott} and a low noise~\cite%
{Qasimi,Teklu,YCLiu,Brivio,Genoni,Genoni12,Escher,Zhong,Bardhan,Feng2014,Vidrighin,YGao}%
, as well as the photon counters with a high detection efficiency~\cite%
{Calkins} and a large enough number resolution~\cite{PLiu}. Most recently,
Liu \textit{et al.}~\cite{PLiu} investigated the influence of the finite
number resolution of the PNRDs in the squeezed-state interferometry and
found that the theoretical precision~\cite{Smerzi08,Liu,Lang} can still be
attainable, provided the resolution threshold $N_{\mathrm{res}}>5\bar{n}$.

In this work, we further investigate the ultimate phase estimation of the
squeezed-vacuum $\otimes $ coherent-state light interferometry using the
PNRDs with a relatively low number resolution $N_{\mathrm{res}}\sim \bar{n}$%
. We first calculate the CFI of a finite-$N$ photon state that post-selected
by the detection events $\{N_{a},N_{b}\}$, with $N_{a}+N_{b}=N$. When the
two light fields are phase matched, i.e., $\cos (\theta_{b}-2\theta _{a})=+1$
for $\theta _{a}=\arg \alpha $ and $\theta _{b}=\arg \xi$, we show that the
CFI of each $N$-photon state equals to that of the QFI. The finite-$N$
photon state under postselection is highly entangled~\cite{Hofmann,Afek},
but cannot improve the estimation precision~\cite{Combes,Pang,Haine}. This
is because the CFI or equivalently the QFI is weighted by the generation
probability of the finite-$N$ photon state, which is usually very small as $%
N\gg1$. To enlarge the CFI and hence the ultimate precision, all $N$-photon
detection events with $N\leq N_{\mathrm{res}}$ have to be taken into
account. We present an analytic solution of the total Fisher information to
show that the Heisenberg scaling of the estimation precision is still
possible even for the PNRDs with $N_{\mathrm{res}}\sim \bar{n}$.

\section{Fisher information of the $N$-photon detection events}

\begin{figure}[tbph]
\begin{centering}
\includegraphics[width=0.6\columnwidth]{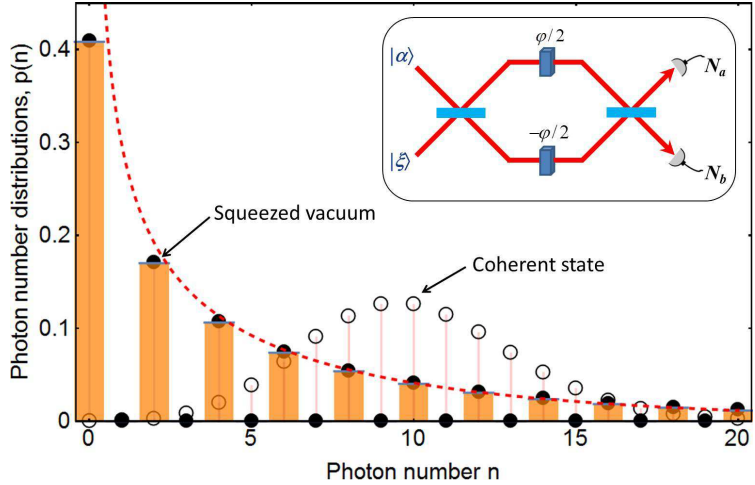}
\caption{Photon number distributions of a coherent state $|\alpha\rangle$ (open circles) and a squeezed vacuum $|\xi\rangle$ (solid circles), with $|\alpha|^2=2\sinh^2|\xi|=10$. The red dashed line is given by Eq.~(\ref{Stirling}). Inset: Photon-counting measurement at the output ports of the MZI that fed by a coherent state and a squeezed vacuum. } \label{fig1}
\end{centering}
\end{figure}

As illustrated schematically by the inset of Fig.~\ref{fig1}, we consider
the Mach-Zehnder interferometer (MZI) fed by a coherent state $|\alpha
\rangle $ and a squeezed vacuum $|\xi \rangle $, i.e., a product input state
$|\psi _{\mathrm{in}}\rangle =|\alpha \rangle _{a}\otimes |\xi \rangle _{b}$%
, where the subscripts $a$ and $b$ denote two input ports (or two orthogonal
polarized modes). Photon-number distributions of the two light fields are
depicted by Fig.~\ref{fig1}, indicating that the squeezed vacuum contains
only even number of photons~\cite{Gerry}, with the photon number
distribution
\begin{equation}
p(2k)=\left\vert s_{2k}\right\vert ^{2}\approx \frac{1}{\cosh |\xi |}\frac{%
(\tanh |\xi |)^{2k}}{\sqrt{\pi k}},  \label{Stirling}
\end{equation}%
where $s_{n}=\langle n|\xi \rangle $ for odd $n$'s are vanishing (see the
Appendix), and we have used Stirling's formula $k!\approx \sqrt{2k\pi }%
(k/e)^{k}$. Furthermore, one can see that the squeezed vacuum shows
relatively wider number distribution than that of the coherent state.

Without any loss and additional reference beams in the paths, we now
investigate the ultimate estimation precision with the $N$-photon detection
events, i.e., all the outcomes $\{N_{a},N_{b}\}$ with $N_{a}+N_{b}=N$, where
$N_{a}\ $and $N_{b}$ are the number of photons detected at the two output
ports. For each a given $N$, it is easy to find that there are ($N+1$)
outcomes as $\mu \equiv (N_{a}-N_{b})/2\in \lbrack -N/2,+N/2]$. To calculate
the CFI of the $N$-photon detection events, we first rewrite the input state
as $|\psi _{\mathrm{in}}\rangle =\sum_{N}\sqrt{G_{N}}|\psi _{N}\rangle $
\cite{Donner}, where $G_{N}$ is the generation probability of a finite $N$%
-photon state:
\begin{equation}
|\psi _{N}\rangle =\frac{1}{\sqrt{G_{N}}}\sum_{k=0}^{N}c_{N-k}(\theta
_{a})s_{k}(\theta _{b})|N-k\rangle _{a}\otimes |k\rangle _{b}.  \label{psiN}
\end{equation}%
Note that the probability amplitudes $c_{n}(\theta _{a})=\langle n|\alpha
\rangle $ and $s_{k}(\theta _{b})=\langle k|\xi \rangle$ depend on the
phases of two light fields $\theta _{a}=\arg \alpha $ and $\theta _{b}=\arg
\xi$ (see Appendix A). In addition, the generation probability is also a
normalization factor of the $N$-photon state and is given by $%
G_{N}=\sum_{k=0}^{N}|c_{N-k}s_{k}|^{2}$.

Next, we assume $|\psi _{N}\rangle $ as the input state of the MZI and
consider the photon-counting measurements over $\exp (-i\varphi J_{y})|\psi
_{N}\rangle $, where $J_{y}=(a^{\dag }b-b^{\dag }a)/(2i)$ and the unitary
operator comes from sequent actions of the first 50:50 beam splitter, the
phase accumulation in the path, and the second 50:50 beam splitter, as
illustrated by the inset of Fig.~\ref{fig1}. According to Refs.~\cite%
{Helstrom,Kay,Braunstein,Luo,Smerzi09,Giovannetti}, the ultimate precision
in estimating $\varphi $ is determined by the CFI:%
\begin{equation}
F_{N}(\varphi )=\sum_{\mu =-J}^{+J}\frac{\left[ \partial P_{N}(\mu |\varphi
)/\partial \varphi \right] ^{2}}{P_{N}(\mu |\varphi )},  \label{FN}
\end{equation}%
where $P_{N}(\mu |\varphi )=|\langle J,\mu |\exp (-i\varphi J_{y})|\psi
_{N}\rangle |^{2}$ denotes the conditional probability for a $N$-photon
detection event. For brevity, we have introduced the Dicke states $|J,\mu
\rangle =|J+\mu \rangle _{a}\otimes |J-\mu \rangle _{b}$, with the total
spin $J=N/2$.

To obtain an explicit form of the CFI, we assume that the two injected
fields are phase matched~\cite{Smerzi08,Liu,Lang}, i.e., $\cos (\theta
_{b}-2\theta _{a})=+1$, for which Eq.~(\ref{psiN}) becomes $|\psi _{N}\rangle =\exp (iN\theta _{a})|\tilde{%
\psi}_{N}\rangle$~\cite{PLiu}. Here, $\theta _{a}$ is an \emph{arbitrary}
phase of the coherent-state light and $|\tilde{\psi}_{N}\rangle $ denotes a
postselected $N$-photon state and is given by Eq.~(\ref{psiN}) for $\theta
_{a}=\theta _{b}=0$. Under this phase-matching condition, the conditional probabilities can be expressed as $P_{N}(\mu |\varphi )=[\langle J,\mu |\exp
(-i\varphi J_{y})|\tilde{\psi}_{N}\rangle ]^{2}$, due to $\langle J,\mu
|\exp (-i\varphi J_{y})|\tilde{\psi}_{N}\rangle \in \mathbb{R}$, which in
turn gives%
\begin{equation*}
\frac{\partial P_{N}(\mu |\varphi )}{\partial \varphi }=2\sqrt{P_{N}(\mu
|\varphi )}\langle J,\mu |(-iJ_{y})e^{-i\varphi J_{y}}|\tilde{\psi}%
_{N}\rangle \in \mathbb{R},
\end{equation*}%
and hence the CFI (see Appendix A):%
\begin{eqnarray}
F_{N}(\varphi )\!\! &=&\!\!4\sum_{\mu =-J}^{+J}\left[ \langle J,\mu
|(-iJ_{y})e^{-i\varphi J_{y}}|\tilde{\psi}_{N}\rangle \right] ^{2}=4\langle
\tilde{\psi}_{N}|J_{y}^{2}|\tilde{\psi}_{N}\rangle  \notag \\
\!\! &=&\!\!\frac{1}{G_{N}}\sum_{k=0}^{N}\left[ N+2k(N-k)+\frac{2k\alpha ^{2}%
}{\tanh \xi }\right] \left( c_{N-k}s_{k}\right) ^{2},  \label{FN2}
\end{eqnarray}%
where we considered the input light fields with the real amplitudes (i.e., $%
\alpha ,\xi \in \mathbb{R}$), so $c_{n}=c_{n}(0)$ and $s_{k}=s_{k}(0)$.
Since $|\tilde{\psi}_{N}\rangle $ contains only even number of photons in
the mode $b$, one can easily obtain $\langle \tilde{\psi}_{N}|J_{y}|\tilde{%
\psi}_{N}\rangle =\mathrm{Im}\langle \tilde{\psi}_{N}|a^{\dag }b|\tilde{\psi}%
_{N}\rangle =0$ and hence the QFI $F_{Q,N}=4\langle \tilde{\psi}%
_{N}|J_{y}^{2}|\tilde{\psi}_{N}\rangle $. Therefore, Eq.~(\ref{FN2})
indicates that the CFI is the same to the QFI of the $N$-photon state $\exp
(-i\varphi J_{y})|\tilde{\psi}_{N}\rangle $. Previously, we have shown that
the CFI or equivalently the QFI can reach the Heisenberg scaling as $%
F_{N}(\varphi )=F_{Q,N}\sim O(N^{2})$~\cite{PLiu}. However, such a quantum
limit is defined with respect to the number of photons being detected $N$,
rather than the injected number of photons $\bar{n}=\alpha ^{2}+\sinh
^{2}\xi $. Furthermore, the $N$-photon state $|\psi _{N}\rangle $ or $|%
\tilde{\psi}_{N}\rangle $ is NOT a real generated state because its
generation probability $G_{N}$ is usually very small, especially when $N\gg
1 $.

\begin{figure}[tbph]
\begin{centering}
\includegraphics[width=0.8\columnwidth]{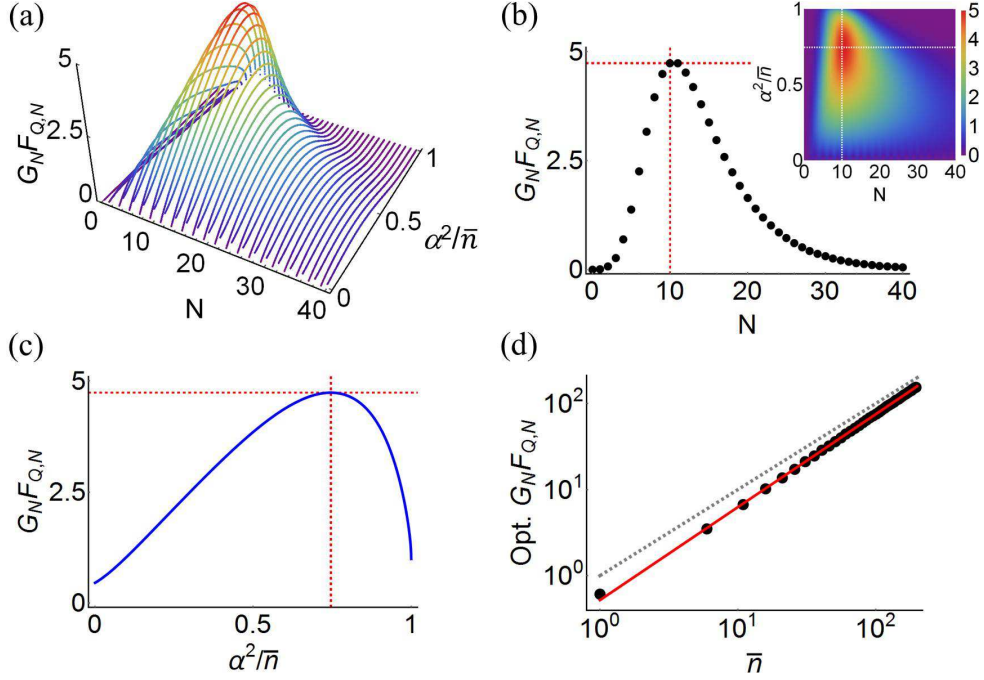}
\caption{For a given $\bar{n}$, the QFI $G_N F_{Q,N}$ reaches its maximum at certain values of $N$ and $\alpha^2/\bar{n}$. (a)-(c) The 3D plot of the QFI for $\bar{n}=8$ and its 2D cross-sections at $\alpha^2/\bar{n}=0.75$ and $N=10$ (marked by the dashed white lines in the inset). (d) Optimal values of the QFI for each a given $\bar{n}\in [1, 200]$, which, fitted by $0.52\bar{n}^{1.08}$ (the red solid), cannot surpass the classical limit (the dashed line) as long as $\bar{n}<10^3$. } \label{fig2}
\end{centering}
\end{figure}

Indeed, the generated state under postselection cannot improve the ultimate
precision for estimating a single parameter~\cite{Combes,Pang,Haine}, since
the CFI is weighted by the generation probability, i.e., $G_{N}F_{N}(\varphi
)$, where $F_{N}(\varphi )=F_{Q,N}$ has been given by Eq.~(\ref{FN2}). As
depicted by Fig.~\ref{fig2}, we find that for a given $\bar{n}=8$, the
weighted CFI or the QFI $G_{N}F_{Q,N}$ reaches its maximum at $N=10$ and $%
\alpha ^{2}/\bar{n}=0.75$. This means that the $10$-photon detection events
give the best precision when the MZI is fed by an optimal input state with $%
\alpha ^{2}=6$ and $\sinh ^{2}\xi =2$. For each a given $\bar{n}\in \lbrack
1,200]$, we optimize $G_{N}F_{Q,N}$ with respect to $\{N,\alpha ^{2}\}$.
From Fig.~\ref{fig2}(d), one can see that the maximum of $%
G_{N}F_{Q,N}$ can be well fitted by $0.52\bar{n}^{1.08}$, which cannot
surpass the classical limit as long as $\bar{n}<10^{3}$. To enlarge the CFI
and hence the ultimate precision, all the detection events have to be taken
into account (see below).

\section{Scaling of the total Fisher information}

Photon counting over a continuous-variable state, there are in general
infinite number of the outcomes and all the $N$-photon detection events $%
\{N_{a},N_{b}\}$ contribute to the CFI. However, the photon number-resolving
detector to data is usually limited by a finite number resolution~\cite%
{Smerzi07,Kardynal}, i.e., $N_{a}+N_{b}=N\leq $ $N_{\mathrm{res}}$, where $%
N_{\mathrm{res}}$ is the upper threshold of a single detector. Taking all
the detectable events into account, the total CFI is given by
\begin{equation}
F(\varphi )=\sum_{N=0}^{N_{\mathrm{res}}}\sum_{\mu =-N/2}^{+N/2}\frac{\left[
\partial P(N,\mu |\varphi )/\partial \varphi \right] ^{2}}{P(N,\mu |\varphi )%
}=\sum_{N=0}^{N_{\mathrm{res}}}G_{N}F_{N}(\varphi ),  \label{CFI1}
\end{equation}%
where $P(N,\mu |\varphi )=|\langle J,\mu |\exp (-i\varphi J_{y})|\psi _{%
\mathrm{in}}\rangle |^{2}$ denote the probabilities for detecting the
photon-counting events $\{N_{a},N_{b}\}$. In the last result, we have
reexpressed the input state as $|\psi _{\mathrm{in}}\rangle =\sum_{N}\sqrt{%
G_{N}}|\psi _{N}\rangle $ and therefore, $P(N,\mu |\varphi )=G_{N}P_{N}(\mu
|\varphi )$, where $G_{N}$ is the generation probability of the $N$-photon
state $|\psi _{N}\rangle $. From Eq.~(\ref{CFI1}), one can easily see that
the total CFI is a sum of each $N$-component contribution weighted by $G_{N}$%
. With only the $N$-photon detection events, the CFI is simply given by $%
G_{N}F_{N}(\varphi )$, as mentioned above.

For the phase-matched input state, we have shown that the CFI of each $N$%
-photon component equals to that of the QFI, i.e., $F_{N}(\varphi )=4\langle
\tilde{\psi}_{N}|J_{y}^{2}|\tilde{\psi}_{N}\rangle =F_{Q,N}$, which in turn
gives%
\begin{equation}
F(\varphi )=\sum_{N=0}^{N_{\mathrm{res}}}G_{N}F_{Q,N}=4\sum_{N=0}^{N_{%
\mathrm{res}}}G_{N}\langle \tilde{\psi}_{N}|J_{y}^{2}|\tilde{\psi}%
_{N}\rangle =F_{Q},  \label{relation}
\end{equation}%
where $F_{Q}$ denotes the total QFI. To see it clearly, let us consider the
QFI\ in the limit of $N_{\mathrm{res}}=\infty $ (i.e., the exact perfect
PNRDs). In this ideal case, the above result becomes
\begin{equation}
F^{\mathrm{(id)}}(\varphi )=4\sum_{N=0}^{\infty }G_{N}\langle \tilde{\psi}%
_{N}|J_{y}^{2}|\tilde{\psi}_{N}\rangle =4\langle \psi _{\mathrm{in}%
}|J_{y}^{2}|\psi _{\mathrm{in}}\rangle =F_{Q}^{\mathrm{(id)}},  \label{ideal}
\end{equation}%
where $F_{Q}^{\mathrm{(id)}}$ is indeed the QFI of the input state $|\psi_{%
\mathrm{in}}\rangle=|\alpha \rangle _{a}\otimes |\xi \rangle_{b}$, for which
$\langle \psi _{\mathrm{in}}|J_{y}|\psi _{\mathrm{in}}\rangle =\mathrm{Im}%
\langle \psi _{\mathrm{in}}|a^{\dag }b|\psi _{\mathrm{in}}\rangle=0$.

It should be pointed out that all the events $\{N_{a},N_{b}\}$ with $%
N_{a}+N_{b}>N_{\mathrm{res}}$ are undetectable and have been discarded in
Eq.~(\ref{CFI1}). However, if we treat them as an additional outcome, the total CFI becomes $F(\varphi )+[\partial P(\mathrm{add}%
|\varphi )/\partial \varphi ]^{2}/P(\mathrm{add}|\varphi )$, where
\begin{equation}
P(\mathrm{add}|\varphi )=1-\sum_{N=0}^{N_{\mathrm{res}}}\sum_{\mu
=-N/2}^{+N/2}P(N,\mu |\varphi ).  \label{Padd}
\end{equation}%
For the perfect MZI considered here, the additional outcome contains no
phase information as $P(\mathrm{add}|\varphi )=1-\sum_{N=0}^{N_{\mathrm{res}%
}}G_{N}$ and hence $\partial P(\mathrm{add}|\varphi )/\partial \varphi=0$.
Therefore, Eq.~(\ref{CFI1}) still works to quantify the ultimate estimation
precision.

Previously, we have considered the photon counters with a large enough
number resolution $N_{\mathrm{res}}$ ($\geq 5\bar{n}$) and found that the
optimal input state contains more coherent light photons than that of the
squeezed vacuum~\cite{PLiu}, rather than the commonly used optimal input
state (i.e., $|\alpha |^{2}\approx \sinh ^{2}|\xi |$). Here, we further
consider the PNRDs with a low resolution threshold $N_{\mathrm{res}}\sim
\bar{n}$. For brevity, we assume the two injected light fields with $\alpha
,\xi \in \mathbb{R}$, for which the phase-matching condition is fulfilled
and hence the CFI still equals to the QFI. Combining Eqs.~(\ref{FN2}) and (%
\ref{CFI1}), we first rewrite the exact result of the QFI (see the Appendix)
as%
\begin{equation}
F_{Q}=\sum_{N_{a}=0}^{N_{\mathrm{res}}}\sum_{N_{b}=0}^{N_{\mathrm{res}%
}-N_{a}}\left[ N_{a}+\left( 1+2N_{a}+\frac{2\alpha ^{2}}{\tanh \xi }\right)
N_{b}\right] \left( c_{N_{a}}s_{N_{b}}\right) ^{2},  \label{Exact}
\end{equation}%
where $c_{n}$ and $s_{k}$ are real, as mentioned above. Next, we note that
the photon number distribution of the squeezed vacuum is usually wider than
that of the coherent state~(see Fig.~\ref{fig1}), so we obtain
\begin{equation*}
\sum_{N_{a}=0}^{N_{\mathrm{res}}}\sum_{N_{b}=0}^{N_{\mathrm{res}%
}-N_{a}}N_{a}f(N_{b})\left( c_{N_{a}}s_{N_{b}}\right) ^{2}\approx
\sum_{N_{a}=0}^{\infty }N_{a}c_{N_{a}}^{2}\sum_{N_{b}=0}^{N_{\mathrm{res}}-%
\bar{n}_{a}}f(N_{b})s_{N_{b}}^{2}=\bar{n}_{a}\sum_{N_{b}=0}^{N_{\mathrm{res}%
}-\bar{n}_{a}}f(N_{b})s_{N_{b}}^{2}.
\end{equation*}%
This is because for a large enough $N_{\mathrm{res}}$, the sum over $N_{a}$
is complete and thereby, $\bar{n}_{a}=\sum_{N_{a}=0}^{\infty
}N_{a}c_{N_{a}}^{2}=\alpha ^{2}$, being average photon number from the input
port $a$. Therefore, we immediately obtain an approximate result of the QFI
\begin{equation}
F_{Q}\approx \bar{n}_{a}\sum_{N_{b}=0}^{N_{\mathrm{res}}-\bar{n}%
_{a}}s_{N_{b}}^{2}+\left( 1+2\bar{n}_{a}+\frac{2\bar{n}_{a}}{\tanh \xi }%
\right) \sum_{N_{b}=0}^{N_{\mathrm{res}}-\bar{n}_{a}}N_{b}s_{N_{b}}^{2}.
\label{appCFI}
\end{equation}%
To validate it, we consider the limit $N_{\mathrm{res}}=\infty $ and a
finite $\bar{n}_{a}$ and obtain
\begin{eqnarray}
F_{Q} &\approx &\bar{n}_{a}+\left( 1+2\bar{n}_{a}+\frac{2\bar{n}_{a}}{\tanh
\xi }\right) \bar{n}_{b}  \notag \\
&=&\bar{n}+2\bar{n}_{a}\bar{n}_{b}\left( 1+\sqrt{1+\frac{1}{\bar{n}_{b}}}%
\right) =F_{Q}^{\mathrm{(id)}},  \label{QFIid}
\end{eqnarray}%
where $\bar{n}_{b}=\sum_{N_{b}=0}^{\infty }N_{b}s_{N_{b}}^{2}=\sinh ^{2}(\xi
)$ is the mean photon number from the port $b$, and $1/\tanh \xi =\sqrt{1+1/%
\bar{n}_{b}}$. Using the relation $2\bar{n}_{b}(1+\sqrt{1+1/\bar{n}_{b}}%
)=\exp (2\xi )-1$, we further obtain the ideal result of the QFI $F_{Q}^{%
\mathrm{(id)}}=\alpha ^{2}\exp (2\xi )+\sinh ^{2}(\xi )$, in agreement with
previous result~\cite{Smerzi08}. When the two input fields are phase matched
and are optimally chosen (i.e., $\bar{n}_{a}\approx \bar{n}_{b}\approx \bar{n%
}/2$)~\cite{Smerzi08,Liu,Lang}, it has been shown that $F_{Q}^{\mathrm{(id)}}
$ can reach the Heisenberg scaling $\sim O(\bar{n}^{2})$. To saturate it,
the exactly perfect PNRDs are needed in the photon counting measurements~%
\cite{Seshadreesan}.

For the imperfect PNRDs with a finite number resolution, we now calculate
analytical result of the QFI. To this end, we first simplify Eq. (\ref%
{appCFI}) as
\begin{equation}
F_{Q}\approx F_{Q}^{\mathrm{(id)}}\left( 1-\frac{1}{\bar{n}_{b}}%
\sum_{N_{b}=N_{\mathrm{res}}-\bar{n}_{a}+1}^{\infty
}N_{b}s_{N_{b}}^{2}\right) ,  \label{appCFI2}
\end{equation}%
where we have used the completeness of $|\xi \rangle $, the relation $(1+2%
\bar{n}_{a}+2\bar{n}_{a}/\tanh \xi )=(F_{Q}^{\mathrm{(id)}}-\bar{n}_{a})/%
\bar{n}_{b}$, and neglected the terms $\sim O(\bar{n}_{a})$. Next, we use
Stirling's formula and replace the sum by an integral, namely
\begin{eqnarray}
F_{Q}\!\! &\approx &\!\!F_{Q}^{\mathrm{(id)}}\left[ 1-\frac{1}{\bar{n}_{b}}%
\int_{N_{\mathrm{res}}-\bar{n}_{a}+1}^{\infty }\frac{N_{b}}{2}p(N_{b})dN_{b}%
\right]  \notag \\
\!\! &=&\!\!F_{Q}^{\mathrm{(id)}}\left[ 1-\sqrt{\frac{\bar{n}_{b}}{1+\bar{n}%
_{b}}}\frac{1}{(\bar{n}_{b}B)^{3/2}}\left( \mathrm{erfc}\left( A\right) +%
\frac{2A}{\sqrt{\pi }}e^{-A^{2}}\right) \right] ,  \label{apprQFI}
\end{eqnarray}%
where, in the first step, $p(N_{b})$ denotes the photon number distribution
of the squeezed vacuum, which can be well approximated by Eq.~(\ref{Stirling}%
). In the second step, $\mathrm{erfc}(A)$ denotes a complementary error
function, $B\equiv \log (1+1/\bar{n}_{b})$, and%
\begin{equation}
A\equiv \sqrt{\frac{N_{\mathrm{res}}-\bar{n}_{a}+1}{2}B}.  \label{A}
\end{equation}%
Clearly, the total QFI depends upon three variables $\{N_{\mathrm{res}},\bar{%
n}_{a},\bar{n}_{b}\}$. For given values of $N_{\mathrm{res}}$ and $\bar{n}$,
one can maximize $F_{Q}$ with respect to $\bar{n}_{a}$ (or $\bar{n}_{b}$) to
obtain the optimal input state and the maximum of the QFI. For instance, let
us consider the limit $N_{\mathrm{res}}\rightarrow \infty $ and hence $%
A\rightarrow \infty $, for which both $\mathrm{erfc}(A)$ and $A\exp (-A^{2})$
are vanishing. Therefore, we immediately obtain the ideal result of the QFI.
The optimal input state can be obtained by maximizing Eq.~(\ref{QFIid}),
which can be approximated as
\begin{equation*}
F_{Q}^{\mathrm{(id)}}\approx \bar{n}+\bar{n}_{a}\left( 4\bar{n}_{b}+1\right)
,
\end{equation*}%
where $\sqrt{1+1/\bar{n}_{b}}\approx 1+1/(2\bar{n}_{b})$ as $\bar{n}_{b}\gg
1 $. With a constraint on $\bar{n}$ ($\gg 1$), it is easy to find that $%
F_{Q}^{\mathrm{(id)}}$ reaches its maximum $\bar{n}(\bar{n}+3/2)$ at $\bar{n}%
_{b}=\bar{n}/2-1/8$, in agreement with previous results~\cite%
{Smerzi08,Liu,Lang}.

\begin{figure}[hptb]
\begin{centering}
\includegraphics[width=1\columnwidth]{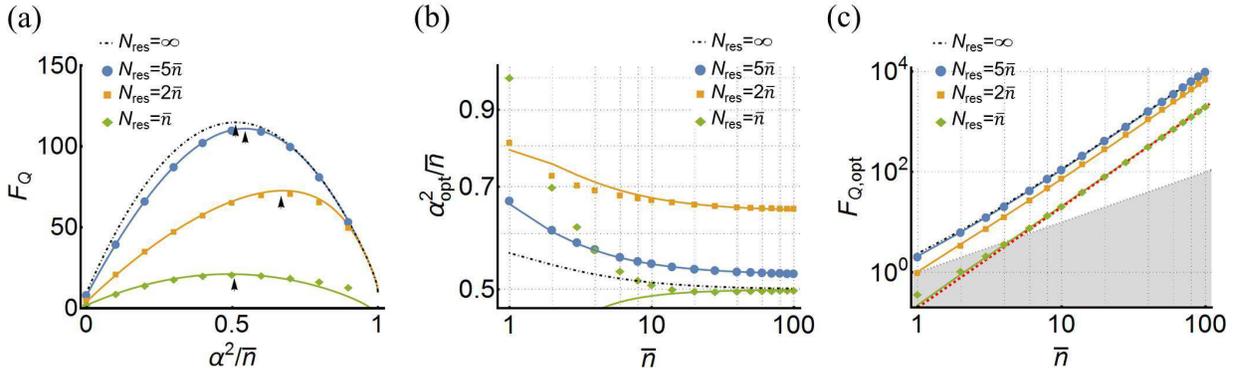}
\caption{For given values of $\bar{n}$ and $N_{\mathrm{res}}$, the total QFI $F_{Q}$ reaches its maximum at a certain value of $\alpha^2/\bar{n}$ (see the arrows). (a) The total QFI $F_{Q}$ as a function of $\alpha^2/\bar{n}$ for $\bar{n}=10$ and $N_{\mathrm{res}}=\bar{n}$ (Diamonds), $2\bar{n}$ (Squares), $5\bar{n}$ (Circles), and $\infty$ (Black dash-dotted line).  (b) and (c) The optimal value of $\alpha^2/\bar{n}$ and the associated QFI $F_{Q,\mathrm{opt}}$ for each a given $\bar{n}\in [1, 100]$, where $N_{\mathrm{res}}$ are chosen the same with (a). The solid lines are analytical results that obtained from Eq.~(\ref{apprQFI}). The red dashed line in (c) is given by Eq.~(\ref{appanaQFI}). The shaded area in (c): A region that below the classical limit $F_Q=\bar{n}$. } \label{fig3}
\end{centering}
\end{figure}

Numerically, the optimal input state can be determined by maximizing Eqs.~(%
\ref{Exact}) and (\ref{QFIid}) with respect to $\alpha ^{2}$ (i.e., $\bar{n}%
_{a}$) for given $\bar{n}$ and $N_{\mathrm{res}}$. As depicted in Fig.~\ref%
{fig3}(a), we choose a fixed mean photon number $\bar{n}=10$ and $N_{\mathrm{%
res}}=\bar{n}$ (the diamonds), $2\bar{n}$ (the squares), $5\bar{n}$ (the
circles), and $\infty $ (the dash-dotted line). The solid lines are obtained from Eq.~(\ref{apprQFI}), which works well to predict the
optimal value of $\alpha ^{2}$, denoted hereinafter by $\alpha _{\mathrm{opt%
}}^{2}$ (see the arrows). In Fig.~\ref{fig3}(b) and (c), we plot $\alpha _{%
\mathrm{opt}}^{2}/\bar{n}$ and $F_{Q,\mathrm{opt}}=F_{Q}(\alpha _{\mathrm{opt%
}}^{2},N_{\mathrm{res}})$ for each a given value of $\bar{n}\in \lbrack
1,100]$, where the values of $N_{\mathrm{res}}$ are taken the same to Fig.~%
\ref{fig3}(a). When $N_{\mathrm{res}}>\bar{n}\gg 1$, the analytical results
of $\alpha _{\mathrm{opt}}^{2}/\bar{n}$ (the solid lines) show good
agreement with the numerical results.

In Figure~\ref{fig3}(c), one can see that $F_{Q,\mathrm{opt}}$ scales as $%
\bar{n}^{2}$ even for the photon counters with a relatively small number
resolution (e.g., $N_{\mathrm{res}}\sim \bar{n}$). To confirm it, we assume
the upper threshold of the number resolution $N_{\mathrm{res}}=\bar{n}$ with
integer $\bar{n}$'s, and calculate analytical result of $F_{Q,\mathrm{opt}}$%
. As shown in Fig.~\ref{fig3}(b), the maximum of the QFI appears at $%
\alpha _{\mathrm{opt}}^{2}/\bar{n}\rightarrow 1/2$ as $N_{\mathrm{res}}=\bar{%
n}\gg 1$, indicating that the optimal input state is the same to the ideal
case (i.e., $\bar{n}_{a}\approx \bar{n}_{b}\approx \bar{n}/2$). Inserting $%
N_{\mathrm{res}}=\bar{n}$ and $\bar{n}_{a}=\bar{n}-\bar{n}_{b}$ into Eq.~(%
\ref{apprQFI}), one can note that the QFI is a function of $\bar{n}_{b}$ for
each a given $\bar{n}$. Therefore, the term $\mathrm{erfc}\left( A\right) $
can be expanded in series of $1/\bar{n}_{b}$,
\begin{equation}
\mathrm{erfc}\left( A\right) =\mathrm{erfc}\left( \frac{1}{\sqrt{2}}\right) -%
\frac{\bar{n}_{b}^{-1}}{2\sqrt{2e\pi }}+O(\bar{n}_{b}^{-2}),\text{ \ \ \ }
\end{equation}%
and similarly,%
\begin{equation}
\frac{2A}{\sqrt{\pi }}e^{-A^{2}}=\sqrt{\frac{2}{e\pi }}-\frac{\bar{n}%
_{b}^{-2}}{8\sqrt{2e\pi }}+O(\bar{n}_{b}^{-3}).
\end{equation}%
When $\bar{n}_{b}\gg 1$, only the leading term dominates in the above
results, and $\bar{n}_{b}B\approx 1$, so we obtain
\begin{equation}
F_{Q}\approx F_{Q}^{\mathrm{(id)}}\left[ 1-\mathrm{erfc}\left( \frac{1}{%
\sqrt{2}}\right) -\sqrt{\frac{2}{e\pi }}\right] \approx 0.2\bar{n}^{2},
\label{appanaQFI}
\end{equation}%
where $F_{Q}^{\mathrm{(id)}}\approx \bar{n}^{2}$ at $\bar{n}_{b}\approx \bar{%
n}/2$, as mentioned above. This scaling shows a good agreement with the numerical result (the diamonds); see Fig.~\ref{fig3}(c). Furthermore,
one can see that the estimation precision can surpass the classical
limit as long as $N_{\mathrm{res}}=\bar{n}>10$.

Finally, it should be mentioned that the Heisenberg limit of phase sensitivity is also attainable using coherent $\otimes$ Fock state as the input~\cite{Pezze}, and a product of
two squeezed-vacuum states~\cite{Lang2}. To achieve such a estimation precision, we show here that it is also important to consider the influence of
a finite number resolution of photon-counting detectors.

\section{Conclusion}

In summary, we have investigated the role of number-resolution-limited
photon counters in the squeezed-state interferometer. Purely with a finite-$%
N $ detection events, we find that the CFI equals to the QFI and is weighted
by the generation probability of the $N$-photon states under postselection.
We numerically show that the maximum of the CFI or equivalently the QFI can
be well fitted as $0.52\bar{n}^{1.08}$, which is slightly worse than the
classical limit as long as $\bar{n}<10^3$. The ultimate precision can be
improved if all the $N$-photon detection events are taken into account. For
the PNRDs with a finite number resolution, the QFI is a sum of different $N$%
-photon components with $N\leq N_{\mathrm{res}}$, which can be approximated
by a simple formula. When $N_{\mathrm{res}}\sim \bar{n}$, our analytical
result shows that maximum of the total QFI scales as $0.2\bar{n}^2$,
indicating that the optimal estimation precision can beat the classical limit for
large enough $\bar{n}$.

\begin{acknowledgments}
We would like to thank Professor H. F. Hofmann for kindly response to our
questions. This work has been supported by the Major Research Plan of the
NSFC (Grant No. 91636108).
\end{acknowledgments}

\appendix

\section{The Fisher information under the phase-matching condition}

We first consider the two light fields with real amplitudes (i.e., $\alpha $%
, $\xi \in \mathbb{R}$), and calculate the QFI of the $N$-photon state under
postselction. In Fock basis, it is given by Eq.~(\ref{psiN}) for the phases $%
\theta _{a}=\theta _{b}=0$,
\begin{equation}
|\tilde{\psi}_{N}\rangle =\frac{1}{\sqrt{G_{N}}}%
\sum_{k=0}^{N}c_{N-k}(0)s_{k}(0)|N-k\rangle _{a}\otimes |k\rangle _{b},
\end{equation}%
where the subscripts $a$ and $b$ represent two input ports or two
orthogonally polarized light modes. The probability amplitudes of the two
fields are given by
\begin{equation}
c_{n}(\theta _{a})\equiv \langle n|\alpha \rangle _{a}=e^{-|\alpha |^{2}/2}%
\frac{|\alpha |^{n}e^{in\theta _{a}}}{\sqrt{n!}},
\end{equation}%
and%
\begin{equation}
s_{k}(\theta _{b})\equiv \langle k|\xi \rangle _{b}=\frac{H_{k}(0)}{\sqrt{%
k!\cosh |\xi |}}\left( e^{i\theta _{b}}\frac{\tanh |\xi |}{2}\right) ^{k/2},
\label{ss}
\end{equation}%
where $H_{2k}(0)=(-1)^{k}(2k)!/k!$ and $H_{2k+1}(0)=0$, are the Hermite
polynomials $H_{k}(x)$ at $x=0$.

Next, we treat $|\tilde{\psi}_{N}\rangle $ as the input state and calculate
the QFI of the output $\exp (-i\varphi J_{y})|\tilde{\psi}_{N}\rangle $. For
the pure state, the QFI is simply given by $F_{Q,N}=4(\langle \tilde{\psi}%
_{N}|J_{y}^{2}|\tilde{\psi}_{N}\rangle -\langle \tilde{\psi}_{N}|J_{y}|%
\tilde{\psi}_{N}\rangle ^{2})$~\cite{Braunstein,Luo,Smerzi09,Giovannetti},
where $J_{y}=(a^{\dag }b-b^{\dag }a)/(2i)$ and $\langle \tilde{\psi}%
_{N}|J_{y}|\tilde{\psi}_{N}\rangle=0$, since $|\tilde{\psi}_{N}\rangle$
contains only even number of photons in the mode $b$. Therefore, we obtain
\begin{eqnarray}
F_{Q,N} =4\langle \tilde{\psi}_{N}|J_{y}^{2}|\tilde{\psi}_{N}\rangle
=\langle (2a^{\dag }ab^{\dag }b+a^{\dag }a+b^{\dag }b)\rangle -\langle
(a^{\dag 2}b^{2}+H.c.)\rangle,  \label{QFIN}
\end{eqnarray}%
where $H.c.$ denotes the Hermitian conjugate and the expectation values are
taken with respect to $|\tilde{\psi}_{N}\rangle $. It is easy to obtain the
first term of Eq.~(\ref{QFIN}),
\begin{eqnarray}
\langle (2a^{\dag }ab^{\dag }b+a^{\dag }a+b^{\dag }b)\rangle =\frac{1}{G_{N}}%
\sum_{k=0}^{N}\left[ 2\left( N-k\right) k+N\right] \left(
c_{N-k}s_{k}\right) ^{2}.  \label{term1}
\end{eqnarray}%
The second term of Eq.~(\ref{QFIN}) can be obtained by calculating%
\begin{eqnarray}
\langle a^{\dag 2}b^{2}\rangle=\frac{1}{G_{N}}%
\sum_{k=2}^{N}c_{N-k+2}s_{k-2}c_{N-k}s_{k} \sqrt{k\left( k-1\right) \left(
N-k+1\right) \left( N-k+2\right) },
\end{eqnarray}%
which is real. Using the relations%
\begin{eqnarray*}
c_{N-k+2} &=&c_{N-k}\frac{\alpha ^{2}}{\sqrt{(N-k+2)(N-k+1)}}, \\
s_{k-2} &=&s_{k}\frac{k}{\sqrt{k(k-1)}}\left( -\frac{1}{\tanh \xi }\right) ,
\end{eqnarray*}%
we further obtain%
\begin{equation}
\langle (a^{\dag 2}b^{2}+H.c.)\rangle =-\frac{1}{G_{N}}\sum_{k=0}^{N}\frac{%
2\alpha ^{2}k}{\tanh \xi }\left( c_{N-k}s_{k}\right) ^{2},  \label{term2}
\end{equation}%
where, in the sum over $k$, we artificially include two vanishing terms for $%
k=0$, $1$. Combining Eqs.~(\ref{term1}) and (\ref{term2}), we obtain the QFI
of the $N$-photon state under the postselection; see Eq.~(\ref{FN2}) in main
text.

Finally, one can note that the above results hold for the two light field
with the complex amplitudes $\alpha $, $\xi $, provided that they are phase
matched, i.e., $\cos (\theta _{b}-2\theta _{a})=+1$. Under this condition,
the $N$-photon state can be expressed as $|\psi _{N}\rangle =\exp (iN\theta
_{a})|\tilde{\psi}_{N}\rangle $, which $\theta _{a}$ is an arbitrary phase
of the coherent light. Similar to Eq.~(\ref{FN2}), the CFI of each $N$%
-photon state is the same with that of the QFI. Furthermore, from Eqs.~(\ref%
{CFI1}) and (\ref{relation}), one can see that the total CFI (or
equivalently, the QFI) is a sum of each $N$-photon component, so we obtain%
\begin{eqnarray}
F_{Q} =\sum_{N=0}^{N_{\mathrm{res}}}G_{N}F_{Q,N} =\sum_{N=0}^{N_{\mathrm{res}%
}}\sum_{k=0}^{N}\left[ 2\left( N-k\right) k+N+\frac{2\alpha ^{2}k}{\tanh \xi
}\right] \left( c_{N-k}s_{k}\right) ^{2}.
\end{eqnarray}%
Setting $k=N_{b}$ and $N-k=N_{a}$, we further obtain the exact result of the
QFI as Eq.~(\ref{Exact}) in main text.


\end{document}